\documentclass{JINST}
\usepackage{graphicx}
\usepackage{amsmath}

\title{A study of gas contaminants and interaction with materials in RPC closed loop systems}

\author{
S.~Colafranceschi$^{a,b}$\thanks{Corresponding author}, 
R.~Aurilio$^b$,
L.~Benussi$^b$, 
S.~Bianco$^b$, 
L.~Passamonti$^b$, 
D.~Piccolo$^b$, 
D.~Pierluigi$^b$, 
A.~Russo$^b$,
M.~Ferrini$^c$, 
T.~Greci$^c$,
G.~Saviano$^c$,
C.~Vendittozzi$^c$,
M.~Abbrescia$^d$, 
C.~Calabria$^d$,
A.~Colaleo$^d$,
G.~Iaselli$^d$,
M.~Maggi$^d$,
S.~Nuzzo$^d$,
G.~Pugliese$^d$,
P.~Verwilligen$^d$,
A.~Sharma$^a$\\
  \llap{$^a$}Physics~Department,~CERN, Geneva,~Switzerland.\\ 
  \llap{$^b$}Laboratori Nazionali di Frascati dell'INFN, INFN, Frascati, Italy.\\ 
  \llap{$^c$}Laboratori Nazionali di Frascati dell'INFN and Universit\`a degli studi di Roma - La Sapienza, Rome, Italy.\\ 
  \llap{$^d$}Politecnico di Bari, Universit\`a di Bari, and INFN Sezione di Bari, Bari, Italy.\\ 

  E-mail: \email{stefano.colafranceschi@cern.ch}
} 

\abstract{Resistive Plate Counters (RPC) detectors at the Large Hadron Collider (LHC) experiments use gas recirculation systems to cope with large gas mixture volumes and costs. 
In this paper a long-term systematic study about gas purifiers, gas contaminants and detector performance is discussed. The study aims at measuring the lifetime of purifiers with unused and used cartridge material along with contaminants release in the gas system. During the data-taking the response of several RPC double-gap detectors was monitored in order to characterize the correlation between dark currents, filter status and gas contaminants.}

\begin{document}

\section{Introduction}
Resistive Plate Counters\cite{Santonico:1981sc} (RPC) detectors are installed at both the ATLAS (A Toroidal LHC Apparatus)\cite{atlas} and CMS (Compact Muon Solenoid)\cite{Chatrchyan:2008aa} experiments at the LHC (Large Hadron Collider) of CERN, Geneva (Switzerland) to provide triggering and synchronization in both barrel and endcap regions as part of the muon system. 
RPCs use a freon-based gas mixture (typically 95.2\% C$_2$H$_2$F$_4$ - 4.5\% Iso-C$_4$H$_{10}$ - 0.3\% SF$_6$) in a recirculation system call Closed Loop (CL)\cite{gassystem}.
Gas mixture is humidified at the 40\% RH level to balance the ambient humidity that affects the resistivity of the highly hygroscopic bakelite.
The CL was designed to cope with large gas mixture volumes and costs. In the closed loop system industrial filters commercially available are in operation to purify the mixture and to prevent contamination collection that affects the RPC performances.
\par
A systematic study of CL gas purifiers has been carried out from 2008 to 2011 at CERN using RPC chambers exposed to cosmic rays and a scaled-down closed loop gas system equipped with several gas analysis sampling points. Goals of the study\cite{Abbrescia:2006LNF} were to observe the release of contaminants in correlation with the dark current increase in RPC detectors, to measure the purifier lifetime\cite{Benussi:2010yx} with unused material, to observe the presence of pollutants.
In this paper, new preliminary results from the 2011 run are shown which characterize the behavior of used purifiers and study the pattern of dark currents increase in the upstream versus the downstream gaps.

\section{Experimental setup}
\begin{figure}[h]
  \begin{center}
    \resizebox{11.0cm}{!}{\includegraphics{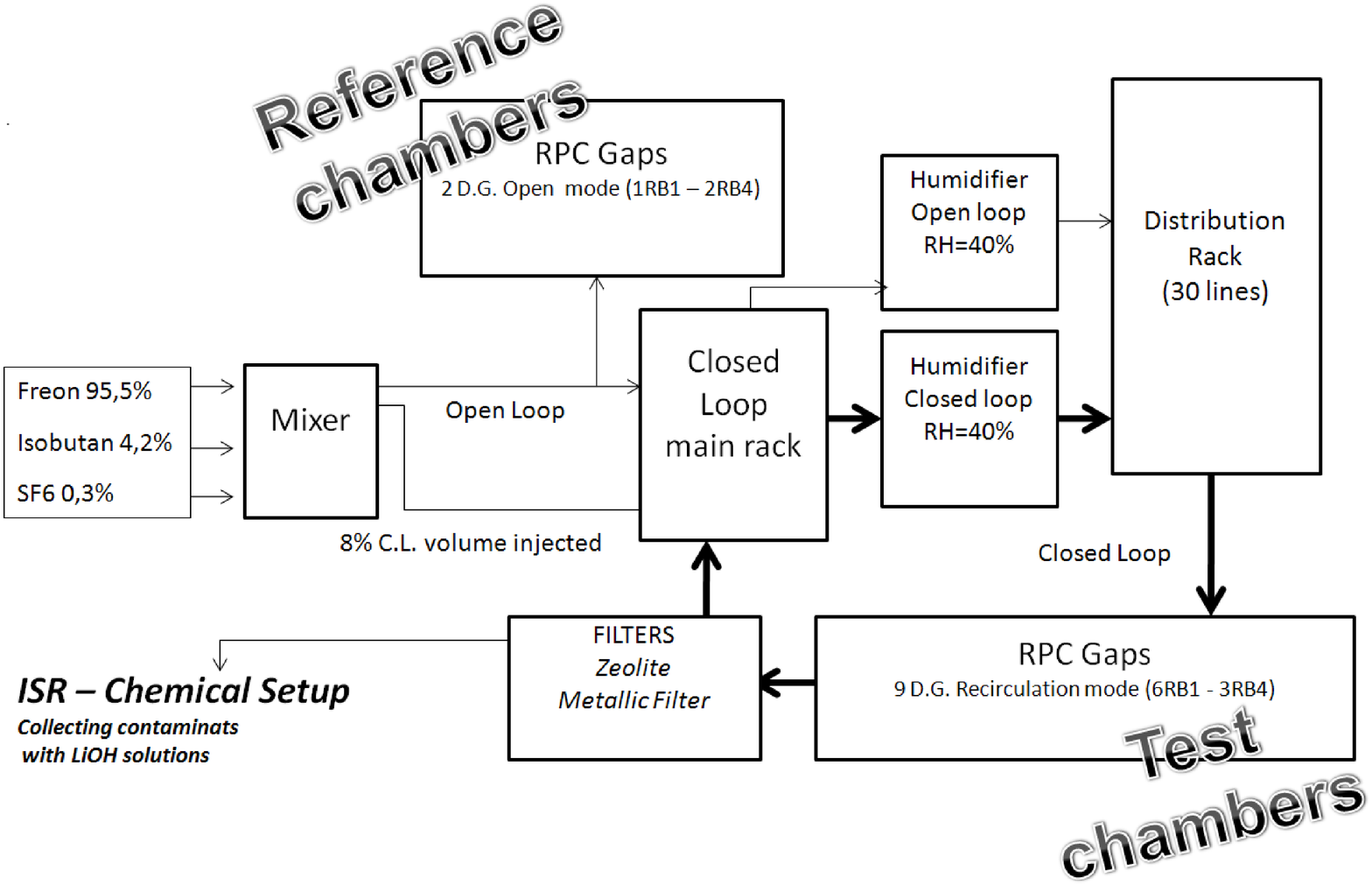}}
    \caption{Schema of the CL setup.}
    \label{FIG:isr_setup}
  \end{center}
\end{figure}

\begin{figure}[h]
  \begin{center}
    \resizebox{9.7cm}{!}{\includegraphics{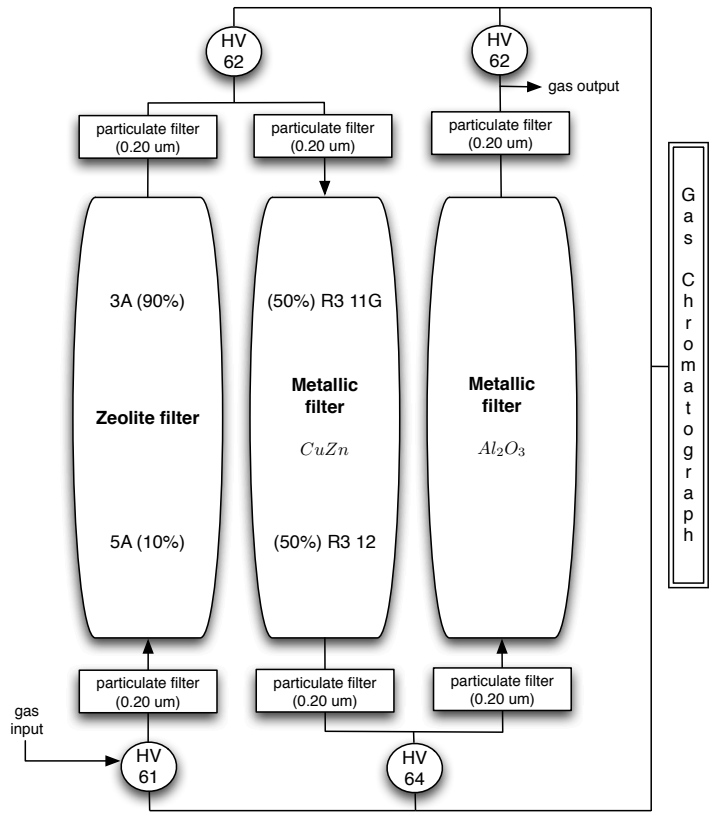}}
    \caption{Purifiers equipped with several sampling points (HV61, 62, 64) before and after each stage.}
    \label{FIG:chem_setup}
  \end{center}
\end{figure}

The experimental setup\cite{Bianco:2009CMSNOTE}\cite{colafranceschi_thesis} is composed of a closed and an open loop gas systems (Fig.~\ref{FIG:isr_setup}). To purify the closed loop gas mixture (that continuously receives 10\% of fresh mixture), commercial filters are used as shown in Fig.~\ref{FIG:chem_setup}. 
\par
The first purifier consists of a 5\AA~~ (10\%) and 3\AA~~ (90\%) type~ zeolite molecular sieve (ZEOCHEM\cite{ZEOCHEM}). The second purifier cartridge is filled with 50\% Cu-Zn filter type R12 (BASF\cite{BASF}) and 50\% Cu filter type R3-11G (BASF) while the third purifier consists of NiAlO$_3$ filter type 6525 (LEUNA\cite{LEUNA}).

\begin{figure}[h]
  \begin{center}
    \resizebox{11.7cm}{!}{\includegraphics{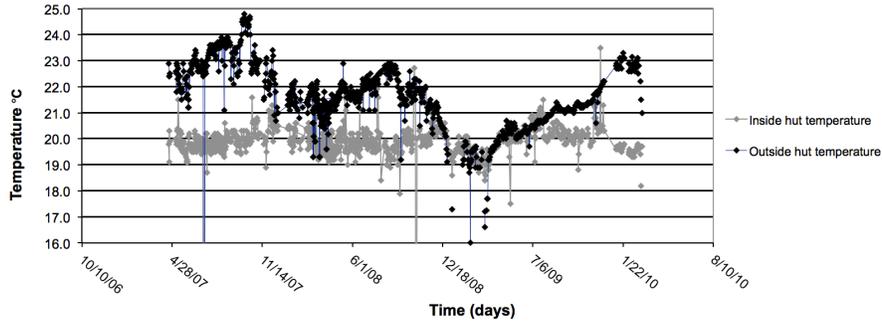}}
    \caption{Temperature trend inside and outside the experimental hut.}
    \label{FIG:isrtemp}
  \end{center}
\end{figure}

\begin{figure}[h]
  \begin{center}
    \resizebox{11.7cm}{!}{\includegraphics{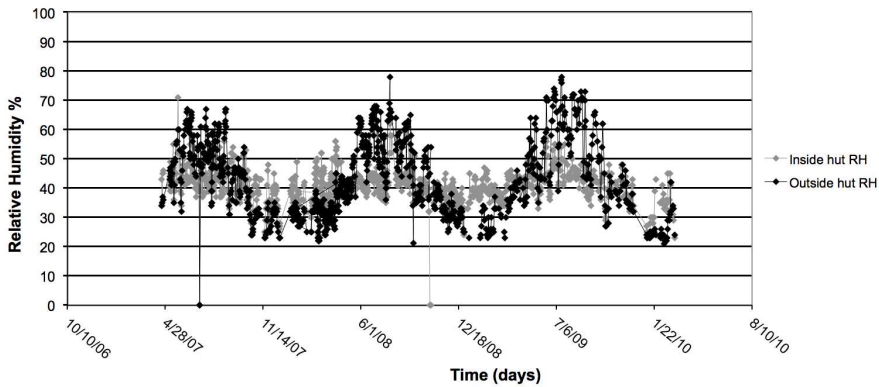}}
    \caption{Relative humidity trend inside and outside the experimental hut.}
    \label{FIG:isrrh}
  \end{center}
\end{figure}

Eleven double-gap RPC detectors are installed in a temperature (Fig.~\ref{FIG:isrtemp}) and humidity (Fig.~\ref{FIG:isrrh}) controlled hut, with online monitoring of environmental parameters. 

\begin{figure}[h]
  \begin{center}
    \resizebox{11cm}{!}{\includegraphics{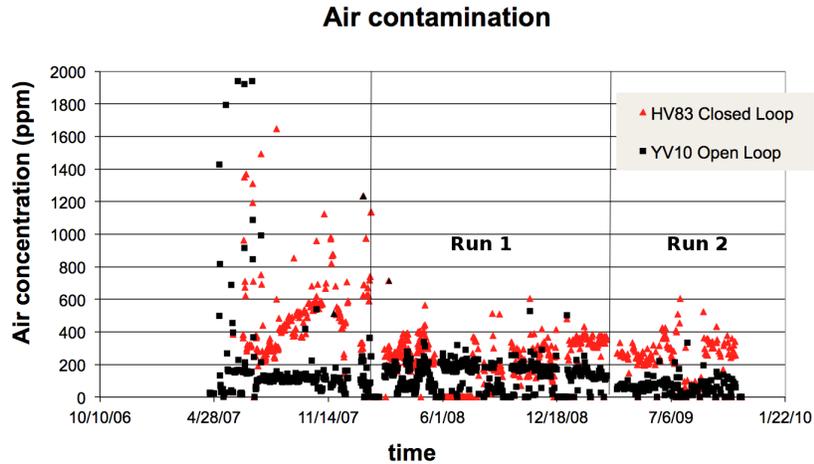}}
    \caption{Air contamination measured in the open and closed loop recirculation system.}
    \label{air_cont}
  \end{center}
\end{figure}

Nine detectors out of eleven are operated in CL mode while two are operated in open loop (OL) mode. Each RPC detector has two gaps (upstream and downstream) whose gas lines are serially connected. The detectors are operated at a 9.2~kV voltage supply. At the working point selected, the anode dark current drawn (due to the high bakelite resistivity) is approximately 1-2~$\mu$A.
Gas sampling points, before and after each filter of the closed loop, allow chemical and gaschromatograph (GC) analysis\cite{Bianco:2009CMSNOTE}. Gas mixture composition is monitored twice a day by GC, which also provides the amount of air contamination, stable over the entire data-taking run and below 300 (100) ppm in closed (open) loop as shown in Fig.~\ref{air_cont}.

\section{Chemical analysis setup}
\begin{figure}[h]
  \begin{center}
    \resizebox{9.0cm}{!}{\includegraphics{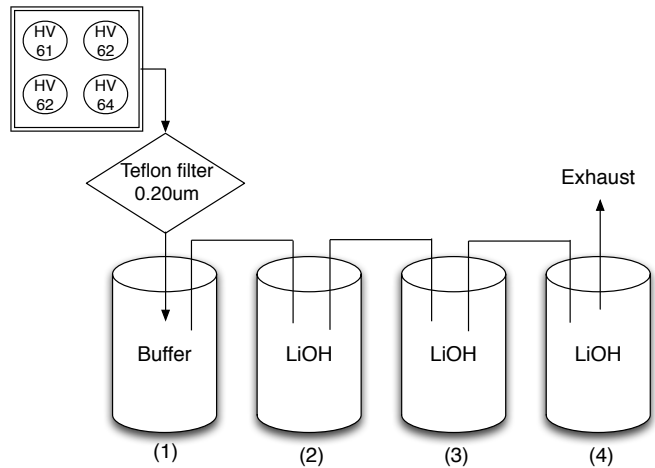}}
    \caption{Chemical analyses are performed using LiOH flasks in which gas is bubbled and contaminants collected.}
    \label{FIG:sampling_point}
  \end{center}
\end{figure}

Chemical analyses have been performed in order to correlate the increase of dark currents with the release of gas contaminants. To identify the contaminants nature, the gas is sampled before and after each purifier, and bubbled into a set of PVC (Polyvinyl chloride) flasks (Fig.~\ref{FIG:sampling_point}).

Flask 1 acts as a buffer to avoid return of LiOH into the CL. Flasks 2, 3 and 4 contained 250~ml solution of LiOH (0.001~mol/l corresponding to 0.024~g/l, optimized to keep the pH of the solution at 11).  The bubbling of gas mixture into the three flasks allows one to capture a wide range of elements that are likely to be released by the system, such as Ca, Na, K, Cu, Zn, Ni, F. At the end of each sampling line the flow is measured to estimate the total amount of gas for the whole period of data-taking.  Sampling points HV61 and HV64 (Fig.~\ref{FIG:sampling_point}) are located before filters (HV61), after zeolite filter (HV62), after Cu/Zn filter (HV64), after Ni filter (HV66).
\par
\begin{figure}[h]
  \begin{center}
    \resizebox{11cm}{!}{\includegraphics{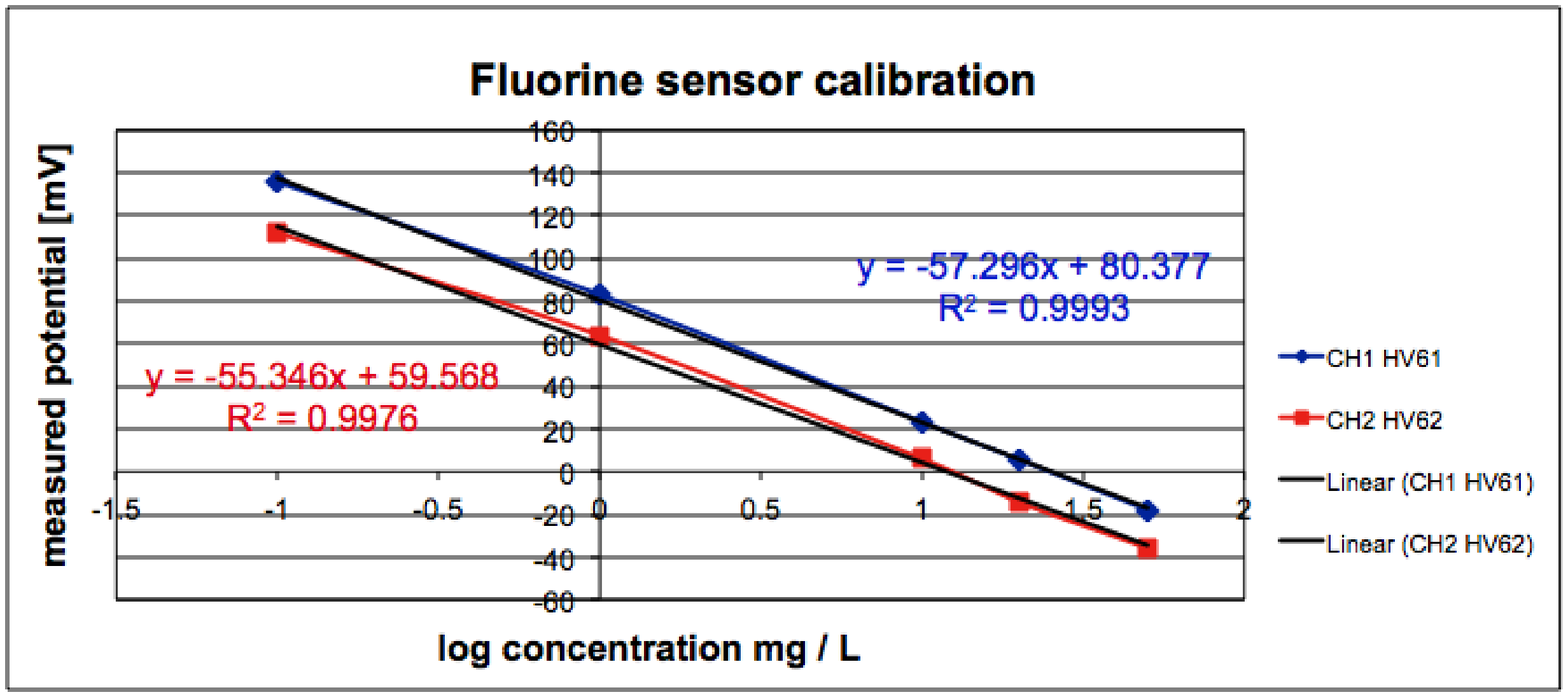}}
    \resizebox{11cm}{!}{\includegraphics{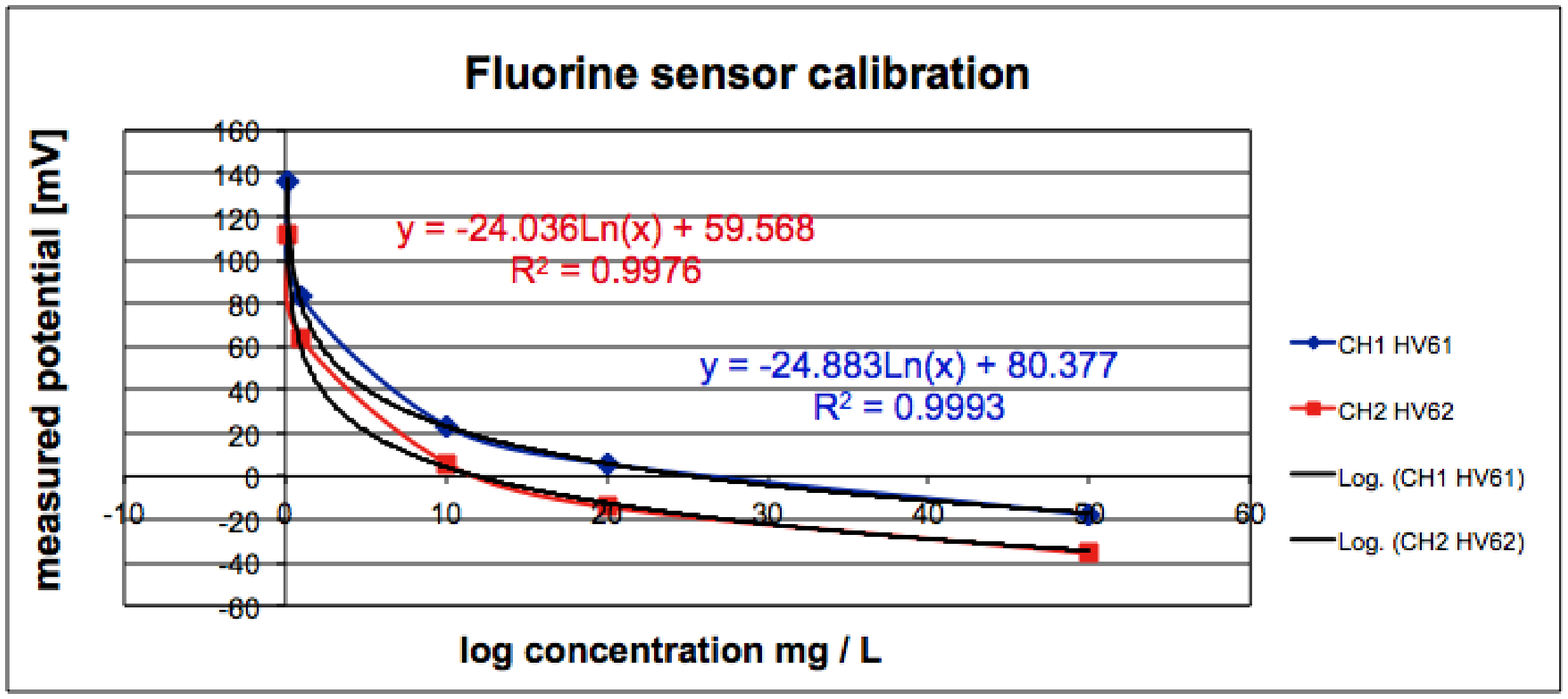}}
    \caption{Calibration curves of fluorine monitoring sensors.}
    \label{fmeno_calib}
  \end{center}
\end{figure}

The fluorine production of RPCs in CL was measured previously in high-radiation condition \cite{Aielli05},\cite{Abbrescia:2006hk}. 
To measure the fluorine production, sampling point HV61 and HV62 are equipped with two additional flaks and fluoride selective electrodes. 
The $\rm{F^-}$ selective electrode adopted\cite{Hanna} is a solid state half-cell sensor that requires, as a separate reference, a silver-silver chloride double junction half-cell reference electrode. Selective electrodes are installed to measure the ionic potential, which is directly connected to the ionic concentration. 
The selective electrodes monitor the collected amount of ions integrated over time by means of a custom software which logs the electrochemical potential every 10 minutes in order to estimate the $\rm{F^-}$ production rate and concentration in the system. 
Both electrodes (reference and sensor) are immersed in a diluted TISAB II solution. This increases and stabilizes the ionic strength of the solution making a linear correlation between the logarithm of the concentration of analyte and the measured potential. 
The selective electrodes were calibrated at the beginning of the run and also during the run itself to double-check a possible shift of the factory settings. Standard solutions containing 0.001~mg/l, 0.005~mg/l, 0.01~mg/l, 0.10~mg/l, 1.0~mg/l, 10.0~mg/l, 20.0~mg/l, 50.0~mg/l concentration of $\rm{F^-}$ are used for absolute calibration. Fig.~\ref{fmeno_calib} shows the calibration curves.

\newpage

\section{Results and discussion}
To describe the operating conditions (Tab.~\ref{summary_data_taking}) the data-taking period is  divided into two runs over three years.
\par
\begin{figure}[htbp]
  \begin{center}
  \resizebox{9cm}{!}{\includegraphics{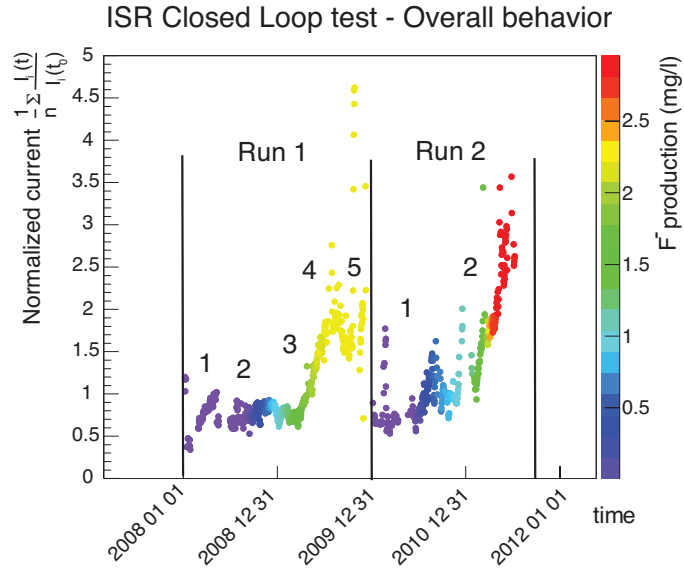}}
   \caption{Upstream gap total current during all runs with F$^{-}$ production.}
    \label{totalcurrentplusfmeno}
  \end{center}
 \end{figure}

\begin{table}[h]
   \caption{Summary table of Closed Loop (CL) and Open Loop (OL) channels.}
    \label{summary_data_taking}
    \begin{center}
      \begin{tabular}{|c|c|c|ccc|} \hline

        Run  & \multicolumn{1}{c|}{Cycle} &
        \multicolumn{1}{c|}{Period} & \multicolumn{3}{c|}{Comment} \\ \hline

          1 & 1 & 29/08/2008 - 11/10/2008 & stable currents & CL & unused
          filters, 9 ch CL, 2 ch OL\\

          1 & 2 & 12/10/2008 - 22/01/2009 & stable currents & CL & unused
          filters 9 ch CL, 2 ch OL\\

          1 & 3 & 23/01/2009 - 28/04/2009 & increasing currents & CL & unused
          filters 9 ch CL, 2 ch OL\\

          1 & 4 & 29/04/2009 - 14/07/2009 & increasing currents & CL & unused
          filters 9 ch CL, 2 ch OL\\

          1 & 5 & 15/07/2009 - 27/07/2010 & decreasing current & OL & used
          filters 0 ch CL, 11 ch OL\\

          2 & 1 & 28/07/2010 - 07/01/2011 & stable currents & CL & used
          filters 7 ch CL, 2 ch OL\\

          2 & 2 & 08/01/2011 - 05/07/2011 & increasing currents & CL & used
          filters 7 ch CL, 2 ch OL\\ \hline

      \end{tabular}
    \end{center}
  \end{table}
Each run is characterized by cycles of operation. Fig.~\ref{totalcurrentplusfmeno} shows the average of all RPC anodic dark currents $I_i(t)$ over  $n$ gaps, normalized by their initial values $I_i(t_0)$. The z-axis scale (color-coded) shows the F$^-$ produced by the system.  The increase of dark anode currents of up-stream gaps in run 1 and run 2 is clearly visible, as well as the increase of F$^-$ concentration. The dark currents of down-stream gaps, as well as the currents of all RPCs in OL, are found stable.
\par
%
% \begin{figure}[htbp]
%  \begin{center}
%  \resizebox{9cm}{!}{\includegraphics{downstreamgap.eps}}
%   \caption{Downstream gap total current during all runs with F$^{-}$ production.}
%    \label{totalcurrentplusfmenodown}
%  \end{center}
% \end{figure}

\begin{figure}[h]
  \begin{center}
    \resizebox{8.3cm}{!}{\includegraphics{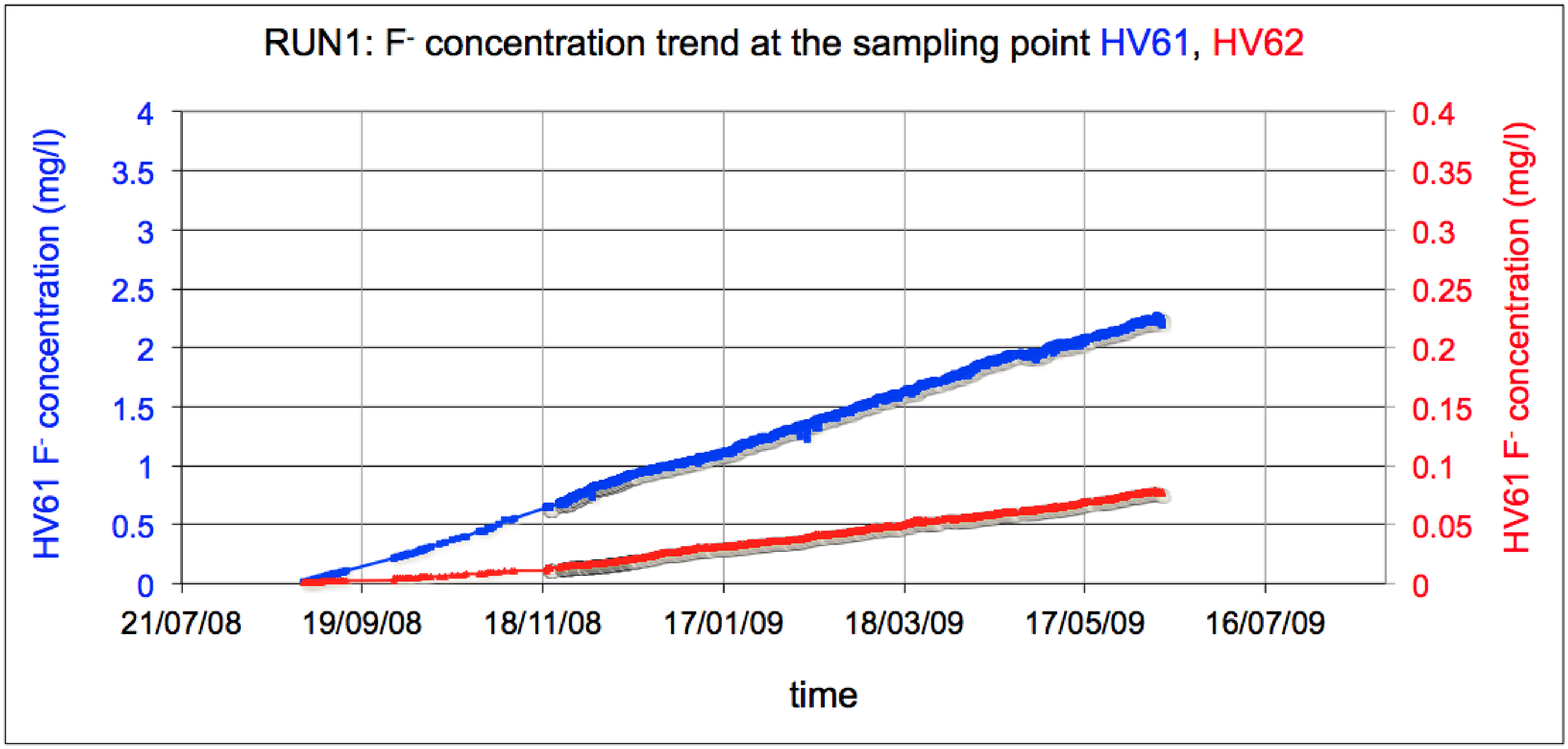}}
    \resizebox{8.3cm}{!}{\includegraphics{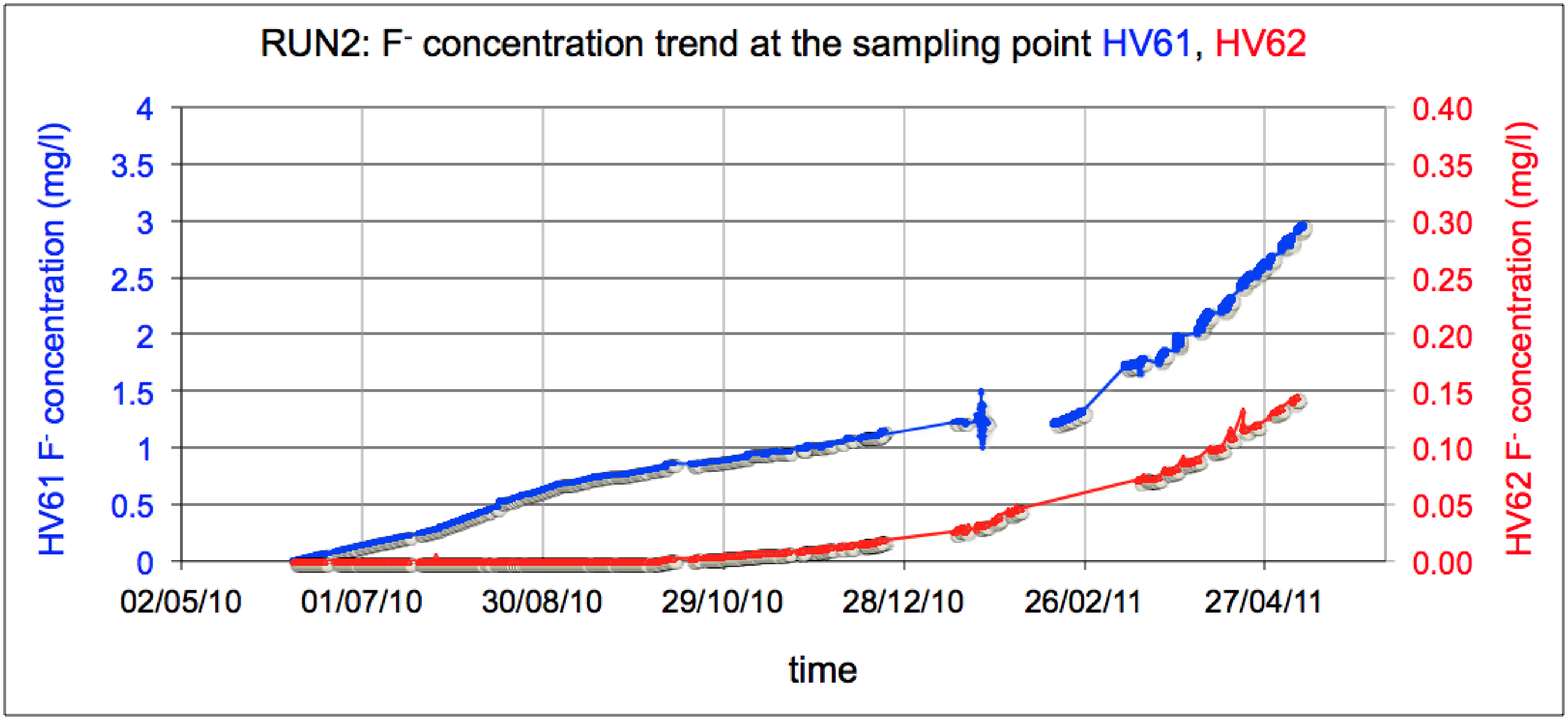}}
    \caption{$\rm{F^{-}}$ production during run 1 (unused filters) and run
    2 (used filters).}
    \label{fmeno_plot_12}
  \end{center}
 \end{figure}

In run 1, the  purifier cartridges are filled with unused material. Eleven double-gap RPC detectors are used, nine in CL and two in OL mode. Cycle 1 and cycle 2 have stable currents up to April 2009, when an increase in the dark current occurs for all up-stream gaps in CL, leaving the down-stream gaps stable. Cycle 4 in particular shows a clear increase of currents, and was terminated before permanent damage occurred to the detector. The lifetime of purifiers is determined by evaluating the duration of cycle 1 and cycle 2
  \begin{equation}
    \tau_{\rm run~1} = 211 \pm 2~{\rm days}
  \end{equation}
The total gas flow is  63$\pm$3~l/h.
We measure the fluorine production (Fig.~\ref{fmeno_plot_12}) during run 1 as 1.10$\pm$0.05~$\rm{\mu mol/l}$ corresponding to a total accumulation in the CL of $(45\pm2)\,10^3~\rm{\mu  mol/l}$.
The purifier lifetime normalized to the  $\rm{F^{-}}$  production is

  \begin{equation}
{\hat\tau}_{\rm run~1} = 4.64\pm0.24~{\rm days/mmol/l}
 \end{equation}

\begin{figure}[htbp]
  \begin{center}
    \resizebox{13.0cm}{!}{\includegraphics{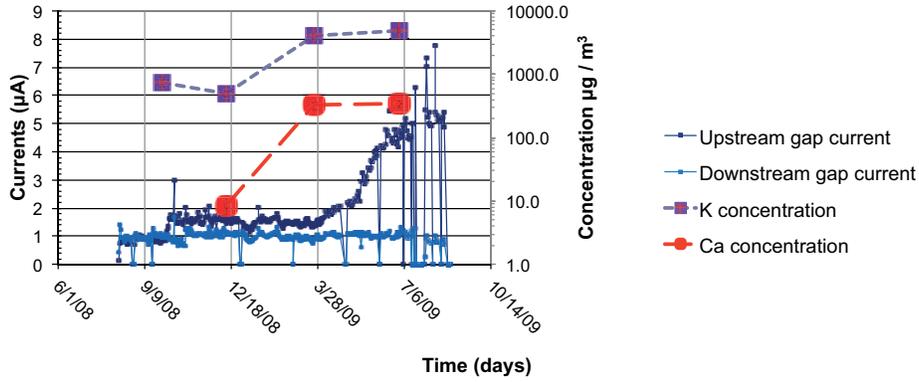}}
    \caption{Dark currents increase (run 1) in the up-stream gap and not in the
    down-stream gap, correlated to the detection of gas contaminants measured
    using the chemical analysis setup.}
    \label{FIG:closedloop}
  \end{center}
 \end{figure}

 Fig.~\ref{FIG:closedloop} shows the typical behaviour of one RPC\index{RPC} detector in closed loop correlated with the concentration of the main contaminants found.
\par
\par
Before starting run 2, all purifiers are regenerated following the CERN gas group standard procedure, i.e. by means of a flushing of hot ($215^{\circ}$C) Ar and H$_2$ mixture (80:20) for twelve hours. 
Nine double-gap RPC detectors are used (seven in CL and two in OL mode).
During
run 2 the flux is measured $\rm{\approx 54\pm3~l/h}$. The currents of down-stream gaps are found stable throughout all cycles as in run 1, while the currents of the up-stream gaps increase. As a cross-check, gas supplies of two gaps of the same RPC detector were swapped to check that in a pair of gaps only the up-stream gap showed currents increase. The lifetime of regenerated purifiers is evaluated:
  \begin{equation}
    \tau_{\rm run~2} = 160 \pm 2~{\rm days}
  \end{equation}
  \par

The $\rm{F^{-}}$ production is measured 0.84$\pm$0.05~$\rm{\mu mol/l}$ (Fig.~\ref{fmeno_plot_12}), corresponding to an accumulation of 33$\pm 2)\,10^3~\rm{\mu  mol/l}$.
The purifier lifetime normalized to the  $\rm{F^{-}}$  production is
  \begin{equation}
    {\hat\tau}_{\rm run~2} = 4.68\pm0.25~{\rm days/mmol/l}
  \end{equation}
 Analyses are in progress in order to confirm the release of contaminants observed in run 1 and shown in Fig.~\ref{FIG:closedloop}.
\par
The lesser $\rm{F^{-}}$ production in run 2 with respect to run 1 is interpreted as due to the smaller number of detectors used in run 2. Although the lifetime of purifiers is measured different in run 1 and run 2, the lifetime normalized to the $\rm{F^{-}}$ production is found compatible within errors, i.e., 4.64$\pm$0.24~days/mmol/l  for run 1 and 4.68$\pm$0.25~days/mmol/l for run 2.
\par
During both run 1 and run 2 the production of $\rm{F^-}$ is efficiently depressed by the zeolite purifier as shown in Fig.~\ref{fmeno_plot_12}. The presence of an excess production of K and Ca in coincidence with the currents increase also suggests a damaging effect of HF\cite{Abbrescia:2006hk}, produced in the system, on the K- and Ca-based zeolite framework. Further analyses are ongoing to verify the presence of contaminants in run~2.

\section{Conclusions}
Preliminary results on studies of contaminants, and on characterizations of materials and gas used in the CL gas system of the CMS RPC muon detector were reported. Quantitative gas chemical analysis were performed by using GC, pH sensors and contaminants detectors. The lifetime of unused purifiers is compatible with the lifetime of regenerated purifiers when normalized to the $\rm{F^-}$  produced in the system. The anodic dark current of up-stream gaps increases when purifiers are exhausted, while the down-stream gaps show stable current.  This behavior is suggestive of a mechanical filtering, i.e., the first gap acts as a filter to the second gap which does not receive a polluted gas mixture. Finally, during run~1 with unused purifiers, traces of K and Ca contaminants were found in correlation with the increase of dark anodic currents. Further studies are in progress in order to ascertain the presence of such contaminants in run~2 with regenerated purifiers.

\end{document}